\documentclass[conference]{IEEEtran}
\IEEEoverridecommandlockouts
\ifCLASSINFOpdf
\else
\fi
 \usepackage[caption=false,font=footnotesize]{subfig}

\usepackage{tikz} % Pacote para desenhar diagramas
\usetikzlibrary{shapes,arrows}
\usetikzlibrary{positioning, arrows.meta, calc, fit}    
\usepackage{float}
\tikzset{
	block/.style = {draw, rectangle, 
		minimum height=0.7cm, 
		minimum width=1.4cm},
	input/.style = {coordinate,node distance=1cm},
	output/.style = {coordinate,node distance=1cm},
	arrow/.style={draw, -latex,node distance=2cm},
	pinstyle/.style = {pin edge={latex-, black,node distance=2cm}},
	sum/.style = {draw, circle, node distance=1cm}
} 

\usepackage{mathtools}
\usepackage{pdfpages}
\usepackage{amsmath}
\usepackage{circuitikz}
\usepackage{multirow}

\begin{document}

% \title{Modeling and Analysis of Coupled Torque Control for Robots Using the Transparent Feedback Structure}

% "Transparency and Stability in Torque/Force Controllers: Quantitative Metrics and Comparative Analysis"

% "Quantitative Analysis of Transparency in Torque/Force Controllers: Metrics, Stability, and Controller Performance"

% "Evaluating Transparency and Coupled Stability in Torque/Force Controllers: An Analytical Approach"

% "Metrics for Transparency and Stability in Torque/Force Controllers: A Comprehensive Study and Model Comparison"

% "Analyzing Transparency and Stability in Torque/Force Controllers: Models, Metrics, and Controller Evaluation"

% "Transparency-Focused Evaluation of Torque/Force Controllers: Analytical Methods and Comparative Insights"

% "Understanding Transparency and Stability in Torque/Force Controllers: A Quantitative and Model-Based Analysis"

% "A Study of Transparency in Torque/Force Controllers: Metrics, Stability, and Comparative Modeling"

% "From Transparency to Stability: Metrics and Models for Torque/Force Controllers in Robotic Systems"

% "Assessing Torque/Force Controllers: Transparency, Stability Metrics, and Analytical Models"

\title{Load-independent Metrics for Benchmarking Force Controllers}

% \title{Method and Metrics for Benchmarking Force/Torque Controllers of Robot Actuators}

%\title{A Novel Approach to Benchmarking Force and Torque Controllers}
% \author{\IEEEauthorblockN{Victor Shime}
% \IEEEauthorblockA{São Carlos School of Engineering\\ University of São Paulo\\
% São Carlos, Brazil\\
% Email: victor.shime@usp.br}
% \and
% \IEEEauthorblockN{Elisa Vergamini}
% \IEEEauthorblockA{São Carlos School of Engineering\\ University of São Paulo\\
% São Carlos, Brazil\\
% Email: elisa.vergamini@usp.br}
% \and
% \IEEEauthorblockN{Thiago Boaventura}
% \IEEEauthorblockA{São Carlos School of Engineering\\ University of São Paulo\\
% São Carlos, Brazil\\
% Email: tboaventura@usp.br }
%
% \IEEEauthorblockN{Andrea Calanca}
% \IEEEauthorblockA{Department of Computer Science\\University of
% Verona\\
% Verona, Italy\\
% Email: andrea.calanca@univr.it }}
\author{
\IEEEauthorblockN{Victor Shime\IEEEauthorrefmark{1},
Elisa G. Vergamini\IEEEauthorrefmark{1},
Cícero Zanette\IEEEauthorrefmark{1},
Leonardo F. dos Santos\IEEEauthorrefmark{1},
Lucca Maitan\IEEEauthorrefmark{1},\\
Andrea Calanca \IEEEauthorrefmark{2}, and
Thiago Boaventura\IEEEauthorrefmark{1}
 \thanks{V. Shime, E. G. Vergamini, C. Zanette, L. F. dos Santos, L. Maitan and 
 T. Boaventura are with University of São Paulo. (emails: {victor.shime, elisa.vergamini, cicero\_zanette, leonardo.felipe.santos, lucca.maitan,  tboaventura\}@usp.br)}
 A. Calanca is with University of Verona (email: andrea.calanca@univr.it)
 }}

% \IEEEauthorblockA{\IEEEauthorrefmark{1}São Carlos School of Engineering, University of São Paulo, São Carlos, Brazil\\ Email: \{victor.shime, elisa.vergamini, cicero\_zanette, leonardo.felipe.santos, lucca.maitan,  tboaventura\}@usp.br}
% \vspace{5pt}
% \IEEEauthorblockA{\IEEEauthorrefmark{2}Department of Computer Science, University of Verona, Verona, Italy\\
% Email: andrea.calanca@univr.it }
}


\maketitle


\begin{abstract}
% Torque-controlled actuators are essential for mechatronic systems that interact closely with their environment, such as legged robots, collaborative manipulators, and exoskeletons. The performance and stability of these actuators are influenced by more than just controller design and robot or actuator dynamics; they also depend significantly on the characteristics of the load (which eventually include interaction with humans or unstructured environments). This load dependence introduces a need for new frameworks to properly assess and compare torque controllers. In this short paper, we use a modeling approach that captures the impact of load on the closed-loop dynamics of torque-controlled systems. Building on this model, we propose methods and quantitative metrics, such as the Passivity Index Interval, which combines passivity and small-gain conditions for a less conservative coupled stability measure than using passivity alone. These metrics can be applied alongside traditional control metrics, such as settling time and bandwidth, to provide a more complete characterization of torque-controlled systems. We validate the effectiveness of our proposed metrics through an experimental comparison of linear actuator force controllers.
%
Torque-controlled actuators are critical components in mechatronic systems that closely interact with their environment, such as legged robots, collaborative manipulators, and exoskeletons. The performance and stability of these actuators depend not only on controller design and system dynamics but also significantly on load characteristics, which may include interactions with humans or unstructured environments. This load dependence highlights the need for frameworks that properly assess and compare torque controllers independent of specific loading conditions. In this short paper, we concisely present a modeling approach that captures the impact of load on the closed-loop dynamics of torque-controlled systems. Based on this model, we propose new methods and quantitative metrics, including the Passivity Index Interval, which blends passivity and small-gain theory to offer a less conservative measure of coupled stability than passivity alone. These metrics can be used alongside traditional control performance indicators, such as settling time and bandwidth, to provide a more comprehensive characterization of torque-controlled systems. We demonstrate the application of the proposed metrics through experimental comparisons of linear actuator force controllers.
\end{abstract}

\section{Introduction and Literature Review}

Torque or force control is commonly employed in mechatronic systems that physically interact with external environments, humans, or payloads. Typical examples include legged robots, collaborative manipulators, and exoskeletons. In such systems, torque tracking performance and stability properties are critical for enabling effective and safe physical interactions. It is well established that these properties depend not only on the robot's characteristics and control algorithms but also on the external environment that physically interacts with the robot, including for instance human users or additional payloads \cite{Calanca2021Enhancing,Calanca2020Actuation,Boaventura2012}.

% In this paper, we consider force-controlled actuators connected through a rotational joint to a link, often modeled as a position-dependent rigid body. This is the common case in robotics, where the controlled actuator drives a rotational joint with position-dependent rigid body model. However, the connected load is often unknown or subject to dynamic changes, as in legged robots, where it may even vary with joint position alone. This variability complicates the design of actuator-controller systems that are both fast and robustly stable under dynamic conditions \cite{Boaventura2012}.

High-level controllers--such as full-body motion planners \cite{Torque_RL,Grandia2019}--typically assume that low-level torque or force controllers behave as ideal torque sources, instantly producing the desired torque at each joint. In practice, however, this assumption is rarely satisfied. The dynamic coupling between the actuator and the connected load impose performance degradation and stability challenges, making sensitivity to load variations a key design consideration for torque-controlled systems.

Several benchmarking methods for torque control have been proposed. Some focus on end-effector force control \cite{Behrens2018, Falco2016, BRUHM2015}, introducing factors beyond actuator torque tracking, while others evaluate performance under a single, locked output-load condition \cite{Ugurlu2022}, typically a locked-output configuration that neglects load dynamics. Robust control approaches model the load as a parametric uncertainty \cite{Niksefat2001}, but their applicability is limited when load variations are highly nonlinear or nonparametric. Broader methods simulate multiple load conditions and employ statistical evaluation metrics \cite{vicario2021benchmarking}; however, they require significant experimental or computational resources, limiting their practical feasibility.

An alternative modeling approach characterizes the closed-loop system via two subsystems: one for torque tracking under blocked-load conditions, and another for apparent impedance when the desired torque is zero \cite{Robinson1999, Roozing2017, Rampeltshammer2023, Shime2023}. This apparent impedance relates to the concept of \emph{transparency}, describing the actuator’s closed-loop backdrivability when disturbed. Although this perspective yields valuable insights into system performance and stability, the lack of objective, standardized metrics for quantifying transparency complicates fair comparisons across controllers. Traditional qualitative assessments—such as inspecting the magnitude of transparency across frequency—are limited in their ability to compare controllers. They provide useful insight only when one controller exhibits uniformly lower transparency than another over the entire frequency spectrum, which is rare in practice. 

Overall, the absence of load-independent benchmarking frameworks limits the ability to fairly compare and directly assess actuator-controller performance and stability \cite{Rampeltshammer2020, Liu2021, Chung2022, Song2023}.
In this short paper, we address this limitation by proposing a set of novel, load-independent analytical metrics that can be universally applied to any force- or torque-controlled actuator system—whether electrically or hydraulically actuated, and whether employing linear or rotational joints.
Specifically, our contributions are twofold:
\begin{itemize}
%\item We introduce a modeling approach that explicitly captures the effects of load variations on closed-loop dynamics;
\item We propose new quantitative metrics, including the \emph{Passivity Index Interval}, that jointly incorporate passivity and small-gain theory, providing a less conservative stability analysis than traditional passivity-only approaches; \item We experimentally validate the proposed metrics by benchmarking the performance of different linear actuator force controllers.
\end{itemize}
By establishing objective, load-independent evaluation methods and metrics, our work aims to advance the standardization of force/torque controller benchmarking, ultimately supporting improved actuator and controller design.

\section{Force control modeling}\label{sec:moddeling}

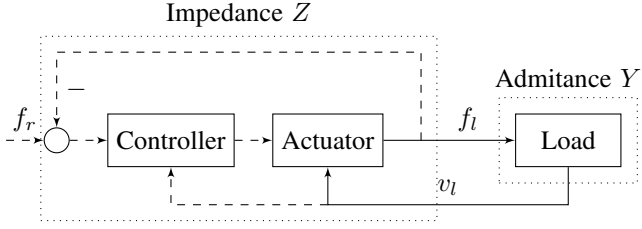
\begin{figure}[tb]
        \centering
        \begin{tikzpicture}[auto, >=latex']
            %Nodes
            \node [block] (controller) {Controller};
            \node [block, right=0.5 of controller,align=center] (actuator) {Actuator};
            \node [sum,left=0.5 of controller] (c1) {};
            \node [coordinate, right=0.5 of actuator] (c2) {};
            \node [block, right=1.25 of c2] (load) {Load};
            \node [coordinate, above=1.0 of c1] (c4) {};
            \node [coordinate, below=0.5 of actuator] (c5) {};
            \node [coordinate, left=0.5 of c1] (c6) {};

            \node[draw, dotted, fit=(controller) (actuator) (c4)(c5) (c2), inner sep=6pt, label={above:Impedance $Z$}] (dottedBox) {};
            \node[draw, dotted, fit=(load) , inner sep=6pt, label={above:Admitance $Y$}] (dottedBox) {};

            %Lines
            \draw[->,dashed] (controller) -- (actuator);
            \draw[->,dashed] (c1) -- (controller);
            \draw[->] (actuator) -- (c2) -- node[pos=0.5, above]{$f_l$}  (load);
            \draw[->, dashed] (c2) |- (c4) -- node[pos=0.5, right]{$-$} (c1);
            \draw[->, dashed] (c5) -| (controller);
            \draw[->,dashed] (c6) -- node[pos=0.5, above]{$f_r$} (c1);
            \draw[->] (load) |-  node[pos=0.75, above]{$v_l$} (c5) -- (actuator);

        \end{tikzpicture}
    \caption{Block diagram for a closed-loop force controller acting on a load through an actuator. Dashed lines represent signals, while continuous lines represent physical connections. The controller and the actuator dynamics can be grouped into an transparency block $Z$, while the load dynamics forms an admittance block $Y$. The feedback of the load velocity $v_l$ from the load into the actuator is intrinsic to physics and always present in force dynamics, while its feedback in the controller is optional and dependent on the controller design. The closed-loop force controller feedbacks the load force $f_l$ and compares it with a reference force $f_r$.}
    \label{fig:openloopIA_interation_model}
\end{figure}

The physical interaction between an actuator-controller combination and a "connected load" (robotic actuator and a generic load) can be modeled as an impedance-admittance coupling, as shown in Fig. \ref{fig:openloopIA_interation_model}.
In this case, the actuator and controller are modeled as an impedance, and the "connected load" as an admittance.

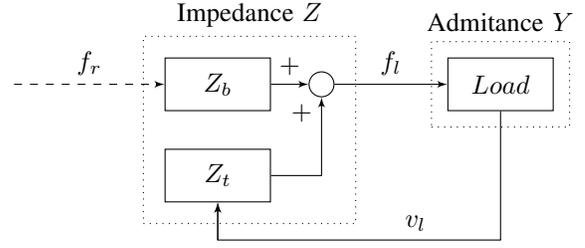
\begin{figure}[tb]
        \centering
        \begin{tikzpicture}[auto, >=latex']
            %Nodes
            \node [block] (actuator_fixed) {$Z_b$};
            \node [block, below=0.5 of actuator_fixed] (transparency) {$Z_t$};
            \node [coordinate, left=2.0 of actuator_fixed] (c1) {};
            \node [coordinate, below=0.5 of transparency] (c2) {};
            \node [sum,right=0.5 of actuator_fixed] (c3) {};
            \node [block, right=1.5 of c3] (load) {$Load$};
            % Dotted rectangle
            \node[draw,dotted, fit=(actuator_fixed) (transparency) (c3), inner sep=8pt, label={above:Impedance $Z$}] (dottedBox) {};
            \node[draw, dotted, fit=(load) , inner sep=6pt, label={above:Admitance $Y$}] (dottedBox) {};

            %Lines
            \draw[->,dashed] (c1) -- node[pos=0.5, above]{$f_r$}  (actuator_fixed);

            \draw[->] (actuator_fixed) -- node[pos=0.5, above]{$+$}  (c3);
            \draw[->] (transparency) -| node[pos=0.9]{$+$}  (c3);
            \draw[->] (load) |-  node[pos=0.65, above]{$v_l$}  (c2) -- (transparency);
            \draw[->] (c2) -- (transparency);
            \draw[->] (c3) --  node[pos=0.5, above]{$f_l$}  (load) ;
        \end{tikzpicture}
    \caption{The closed-loop force-controlled actuator, represented by $Z$, can be divided into two subsystems: blocked ($Z_b$) and transparency ($Z_t$). These two functions are dependent on both the controller and the actuator dynamics.}
    \label{fig:IA_interation_model}
\end{figure}

The actuator block $Z$ in Fig. \ref{fig:openloopIA_interation_model} has 2 inputs, reference force $f_r$ and load velocity $v_l$, and one output, the load force $f_l$.
The proposed methodology is based on splitting $Z$ into 2 subsystems, as depicted in Fig. \ref{fig:IA_interation_model}: $Z_b$, which has $f_r$ as input and represents the force tracking dynamics for a \emph{blocked} load (i.e., $v_l=0$); and $Z_t$, which has $v_l$ as input and represents the residual apparent impedance dynamics when the desired force is zero (i.e., $f_r=0$), that is, the actuator \emph{transparency}. This separation works when the impedance $Z$ is linear. In such a case, the coupled system response to a reference force is defined by the following transfer function:

\begin{equation} \label{eq:forceTF}
    T_y(s)=\frac{F_l(s)}{F_r(s)} = \frac{Z_b(s)}{1-Z_t(s)Y(s)}
\end{equation}

Since feedback controllers tend to dominate open-loop non-linearities, it is reasonable to assume the impedance $Z$, which represents the dynamics of the closed-loop system, as linear. Interestingly, this property usually applies to nonlinear controllers. Some examples are inverse dynamics, and adaptive and sliding-mode controllers, which tend to dominate open-loop nonlinearities or time variance to obtain an approximately linear closed-loop system.

\section{Quantitative Metrics for Transparency}

Regardless of controller architecture, robot kinematics, or actuator choice, load dynamics inherently affect the closed-loop performance and stability of force/torque control. Nevertheless, an appropriate combination of these elements can substantially mitigate that influence.
The load dynamics is coupled to actuator and controller blocks via the physical feedback loop of the load velocity $v_l$, between the load and the transparency block $Z_t$, as shown in Fig. \ref{fig:IA_interation_model}.
This loop is analyzed here using existing feedback control theories. In this section, we propose 4 metrics for assessing the force controller transparency and the impact of the load in the closed-loop response.

\subsection{Load Change Sensitivity (LCS)}

% INTUITIVE DESCRIPTION
Force controllers are typically tuned for a nominal load, and their closed-loop response can vary significantly if the load changes. To quantify the sensitivity to such variations in the response (not necessarily in the stability), we consider the derivative of the closed-loop transfer function \( T_y \) with respect to the load dynamics \( Y \), which captures how small perturbations in \( Y \) affect the system behavior. 

% ASSUMPTION
Since \( T_y \) inherently depends on the load, the sensitivity derivative also reflects this dependency. However, by examining its structure, we can isolate the part governed solely by the actuator-controller pair, independent of specific load realizations. This approach characterizes the intrinsic sensitivity properties of the system.
% DEFINITION
Starting from the system sensitivity with respect to the load dynamics:

\begin{equation}
\frac{d}{dY(s)} \left( \frac{Z_b(s)}{1 - Z_t(s)Y(s)} \right) = -\frac{Z_b(s)Z_t(s)}{(1 - Z_t(s)Y(s))^2} = T_y^2 \frac{Z_t(s)}{Z_b(s)}
\end{equation}

We define the Load-Change Sensitivity (LCS) as:

\begin{equation}
LCS = \max_{\omega \in [0, \omega_b]} \left\| \frac{Z_t}{Z_b} \right\|(\omega) \quad 
\end{equation}

A higher LCS indicates greater sensitivity of the force controller to variations in the load.
% MOTIVATION / PROPERTIES
The LCS captures the worst-case amplification of load-induced perturbations within the relevant frequency range. We restrict the analysis to frequencies up to the locked  output system bandwidth \( \omega_b \) because, beyond that point, \( T_y \) tends to decrease towards zero, altough \( \frac{Z_t}{Z_b} \) may increase, potentially leading to an overestimation of the true sensitivity.  
Moreover, in practice, the bandwidth of the closed-loop system under load is typically lower than that of the locked -output system, reinforcing that the most critical load sensitivity occurs below \( \omega_b \).  
An important implication is that the bandwidth of the blocked system \( Z_b \) must significantly exceed the frequency range where transparency peaks occur; otherwise, the LCS will increase substantially, indicating poor load sensitivity.
% AC: Let's explain this better, possibly using examples to illustrate how the mismatch between the bandwidth and transparency peak affects performance.
% Answer: This is a good idea, but it would take space we don't have.
% \subsection{Transparency Residual (TR)}

% To quantify how transparent the actuator is in closed-loop, we propose the metric, named \emph{Transparency Residual}.

% \begin{equation}
%     TR = ||Z_t||_{2} \quad [N]
% \end{equation}

% As the $H_2$ norm of $Z_t$ is equivalent to the root-mean-square of the impulse response, the TR represents the closed-loop response when the force controller tracks zero force (i.e., $f_r=0$) while an sudden acceleration is applied to the actuator. \emph{Higher} TR implies less transparency.
\subsection{Transparency Residual (TR)}

% INTUITIVE DESCRIPTION
The Transparency Residual (TR) quantifies the force transmitted through the actuator when tracking zero reference force and subjected to external disturbances.
% DEFINITION
The TR is defined as the $H_2$ norm of the transparency transfer function:

\begin{equation}
    TR = \|Z_t\|_2 \quad [\text{N}]
\end{equation}

% MOTIVATION / PROPERTIES
As the $H_2$ norm corresponds to the root-mean-square of the impulse response, the TR represents the residual force response to a unit impulsive disturbance; higher TR values indicate lower transparency.
% NOTE
Although the TR is a standard and well-known metric, it is included here to complete the set of performance indicators for force controller transparency.

\subsection{Passivity Index Interval (PII)}

% INTUITIVE DESCRIPTION
Sector-based stability theorems, such as passivity and small-gain theorems, provide robust methods for ensuring the stability of interacting systems \cite{xia2020sector}. Achieving transparency passivity, characterized by the transparency transfer function \( Z_t \), is a common goal for force controllers, as it promotes stable interaction with passive loads \cite{Calanca2016}, which represent most robotic applications \cite{Dyck}.  
However, passivity constraints are often overly conservative, and as a binary (true-false) condition, passivity alone is insufficient for effectively differentiating controller performance.  
To address this limitation, we partition the frequency domain and apply mixed passivity–small-gain sector arguments: in frequency ranges where the transparency response \( Z_t \) satisfies strictly passive frequency conditions, passivity theory ensures stability; elsewhere, stability is guaranteed by satisfying a small-gain condition.

% COMMENT: AC: the reader here does not get the intuitive meaning of your index and the practical implications. Maybe you can use a figure to illustrate how you combine passivity phase constraints with small gain theorems' gain constraint. Since the gain depends on the load, how can you get rid of it? Also, the system is either passive or non-passive. What does it mean in practice to be "close to passivity"? Standing to your definitions, this applies only to linear systems. Therefore your method applies only to linear controllers? An adaptive controller cannot undergo your benchmarking? Everything must be stated.
% ANSWER: A figure would help, but space constraints may limit it; we improved the explanation in the text by clarifying the frequency partition and the notion that "close to passivity" means \(R_G\) close to 1. Also clarified that the focus is on the passivity properties of the transparency \( Z_t \), not of the controller itself.

% ASSUMPTION
Mixed passivity–small-gain sector conditions \cite{Griggs2007,Griggs2011} relax strict passivity requirements by guaranteeing stability even when the system is not globally passive, provided appropriate conditions are satisfied in each frequency region.  
A system is considered passive in a frequency range if the passivity index \( R_G(\omega) \) satisfies \( R_G(\omega) \leq 1 \) for all \(\omega\) in that range, as defined in Eq.~\eqref{eq:passivity_index}. Strict passivity corresponds to \( R_G(\omega) < 1 \). If \( \max R_G(\omega) > 1 \), the system \( G \) is not passive. For passive systems, \( R_G(\omega) \) typically trends toward \( 1 \) as \( \omega \to 0 \) and \( \omega \to \infty \). 
In regions where strict passivity does not hold, stability is ensured by satisfying the small-gain condition \(\|Z_t(j\omega)\| \cdot \|Y(j\omega)\| < 1\), where \( Y \) denotes the load transfer function.

This approach is particularly relevant for force controllers, since, in systems with stiffness, the load response \( Y \) tends toward zero at sufficiently high frequencies. Consequently, it suffices for \( Z_t \) to exhibit approximate passivity over the relevant frequency interval, while the small-gain condition covers the remaining regions.
% COMMENT: AC: This is not enough. We are working on an omega-passivity concept, but providing the phase is limited within a certain frequency is not sufficient. You need further assumptions on the high-frequency behavior.
% ANSWER: We clarified that we are not claiming full passivity of \( Z_t \), but containment within a mixed sector, and that passive real-world loads generally satisfy the complementary conditions needed to guarantee stability.
% DEFINITION
The system's passivity can be assessed using the passivity index \( R_G(\omega) \) \cite{Anderson1989}, defined as:

\begin{equation}\label{eq:passivity_index}
R_G(\omega) = \| (1-G)(1+G)^{-1} \| (\omega)
\end{equation}

To reflect practical design goals, we propose the Passivity Index Interval (PII), defined as:

\begin{equation}\label{eq:pii}
PII_\epsilon = M(\omega_1, \omega_2)
\end{equation}

\noindent where \( \omega_1 \) and \( \omega_2 \) are the logarithmic frequency bounds within which \( R_{-Z_t}(\omega) \leq 1 - \epsilon \), with \( \epsilon \) a small positive constant, and \( M \) denotes the maximum gain of \( -Z_t \) outside this interval.
\footnote{The negative sign accounts for the assumption of positive feedback in Fig.~\ref{fig:IA_interation_model}, contrasting with the negative feedback convention typically used in classical passivity formulations.}

\begin{figure}
    \centering
    \includegraphics[width=0.9\linewidth]{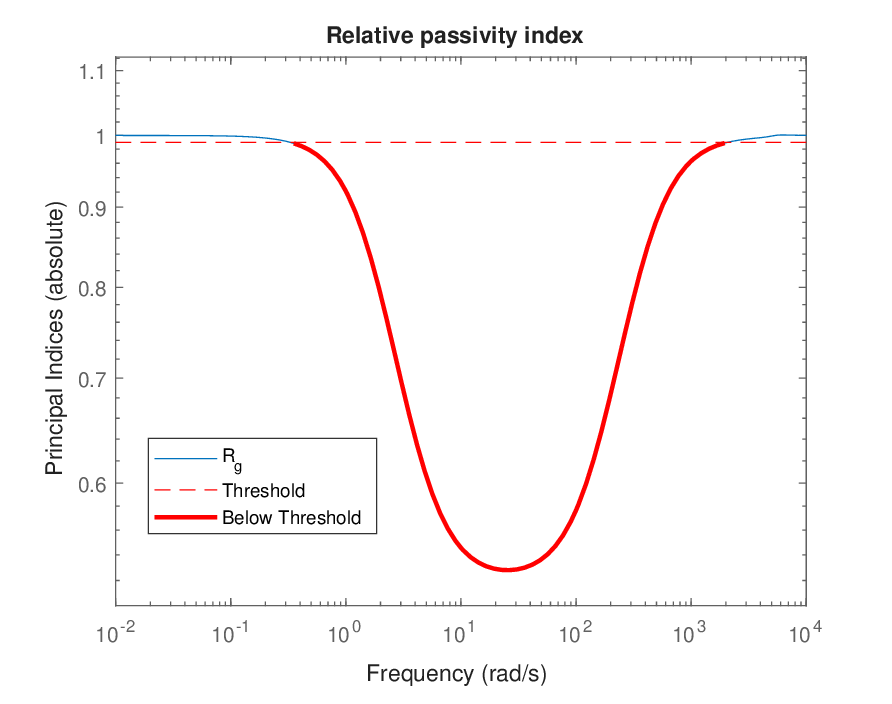}
    \caption{Illustration of the Passivity Index Interval. The thick red segment highlights the frequency range $[\omega_1, \omega_2]$ where the gain $R_g$ is less than $1 - \epsilon$. The $M$ represents the maximum value of $|Z_t|$ outside this interval (thin blue line). According to mixed-sector small gain arguments, if the load is passive and satisfies $|Z_t||Y| < 1$ in the blue line, then the interconnected system remains stable. A smaller $|Z_t|$ allows for larger values of $|Y|$ while still guaranteeing stability.}    \label{fig:PII}
\end{figure}

% MOTIVATION / PROPERTIES
The PII metric captures stability-related performance by quantifying both the frequency interval where strict passivity holds and the maximum deviation elsewhere. Ideally, \( \omega_1 \) and \( M \) should be minimized, while \( \omega_2 \) should be maximized.  
This formulation aligns with practical controller design, where stability must be ensured despite the limited passivity range of real-world actuators interacting with passive or compliant environments.

\subsection{Load Robustness Threshold (LRT)}

% INTUITIVE DESCRIPTION
A key goal in robust control is quantifying the smallest perturbation that destabilizes the system. Classical margins, such as gain and phase, treat amplitude and phase separately. The Load Robustness Threshold (LRT) offers a unified measure by capturing the smallest destabilizing uncertainty in the load dynamics.
% ASSUMPTION
We consider the configuration in Fig.~\ref{fig:IA_interation_model}, where the load \( Y \) is modeled as an uncertain system with norm bounded by one. This normalized formulation is standard in \(\mu\)-analysis, where the actual uncertainty can be written as \( \delta = \alpha \cdot \Delta \) with \( \|\Delta\| \leq 1 \) and \( \alpha \) denoting the relative size.
% COMMENT: The unit-norm assumption allows interpreting \(\mu\) as the inverse of the smallest scaling factor \(\alpha\) such that the scaled uncertainty leads to instability.
% DEFINITION
The LRT is defined as the reciprocal of the peak structured singular value over frequency:

\begin{equation}\label{eq:LRT_def}
LRT = \min_{\omega} \mu^{-1}(\omega) \quad [1]
\end{equation}

\noindent where \( \mu(\omega) \) is the structured singular value at frequency \( \omega \). The LRT thus identifies the smallest gain on the uncertainty such that some admissible perturbation destabilizes the system.

% COMMENT: This guarantees existence of a destabilizing load uncertainty at that threshold, not merely a bound.
% MOTIVATION / PROPERTIES
A higher LRT indicates greater stabilty robustness to load uncertainties, even if system performance may degrade. Unlike traditional margins, the LRT captures simultaneous variation in gain and phase, and explicitly accounts for both the blocked dynamics \( Z_b \) and the transparency transfer function \( Z_t \).  
This provides a more realistic and comprehensive robustness criterion, particularly for systems with strong load–transparency interaction.

\section{Use Case with an Experimental Model}\label{sec:system_analytical}

\begin{figure}[tb]
    \centering
    \includegraphics[scale=0.6]{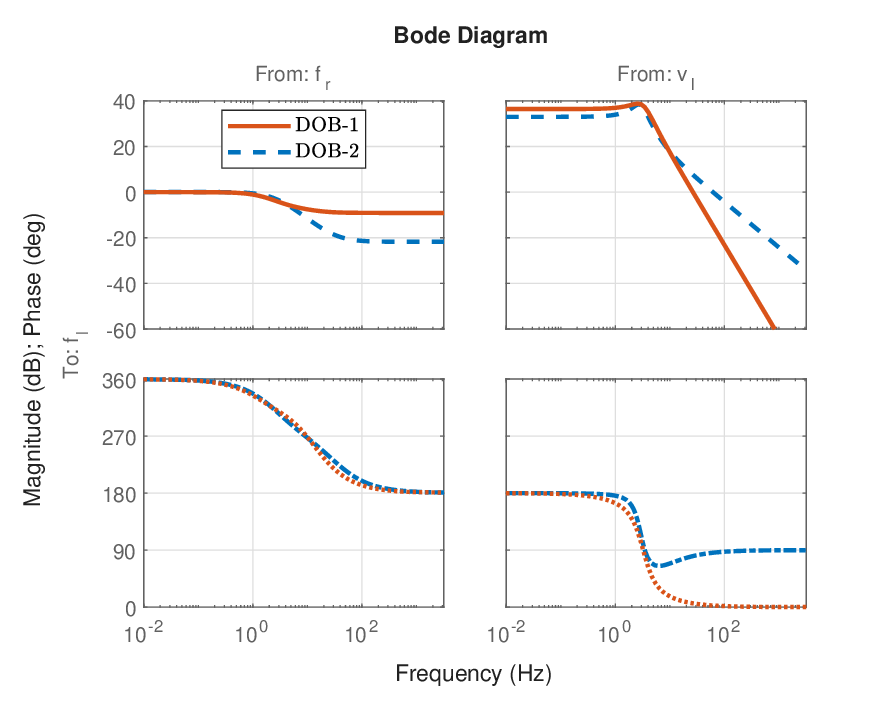}
    \caption{Frequency response of the PMLSM system with two DOB controllers. The plot on the left shows the response of $Z_b$, and the one on the right shows $Z_t$. Controller DOB-2 was tuned for a faster rise time. Transparency initially appears low, increases as the controller attempts to track the load at mid-frequencies, and decreases again at higher frequencies due to spring dynamics.}
    \label{fig:bode}
\end{figure}

In this section, we apply the previously introduced modeling techniques and transparency metrics to evaluate and compare two controller tunings for a permanent magnet linear synchronous motor (PMLSM). Experimental models $Z_b$ and $Z_t$ were identified for each tuning, and the resulting metrics were computed accordingly. The purpose of these experiments is not to optimize controller performance per se, but rather to demonstrate how the proposed metrics reflect changes in controller design. All experiments were carried out on the IC2D test bench~\cite{Vergamini2023ic2d}.
We tested two configurations of Disturbance Observer (DOB)-based controllers~\cite{vergamini2024force}. The first controller, referred to as DOB-1 ($K_p=0.7$, $K_i=0.05$ and $K_d=0.015$), was tuned with increased gain to achieve a reduced rise time. In DOB-2 we used $K_p=0.6$ and kept the other gains. Fig.~\ref{fig:bode} shows the Bode plots of the identified models for both configurations: $Z_b$ on the left and $Z_t$ on the right.

While lower magnitudes of transparency are generally desirable for better interaction, the comparison between the two controllers is not straightforward. DOB-2 exhibits lower transparency at low frequencies, but higher values at high frequencies, suggesting a shift in transparency characteristics rather than a clear improvement.
The transparency metrics in Table~\ref{table:results_a} indicate that DOB-1 achieves better Transparency Ratio (TR), Performance under Input Inversion (PII), and Load Robustness Threshold (LRT), with a slightly higher Load Change Sensitivity (LCS). Complementary results based on $Z_b$ are presented in Table~\ref{table:results_b}, showing that DOB-1 also achieves higher bandwidth and faster rise time.

%\textbf{AC:} This conclusion is extremely rough. Always when we increase gains, we are less sensitive to disturbances!! But this should come with a trade-off in stability margins!! The word "beneficial" does not capture this — it is always a trade-off.

\begin{table}[h]
\caption{$Z_t$ metrics for the experimental data.}
\centering
\begin{tabular}{|c|c|c|c|c|}
\hline
Controller & LCS & TR & PII & LRT \\
\hline
DOB-1 & 121.56 & \textbf{191.98} & \textbf{79.24 (0, 1.247)} & \textbf{0.0119} \\
DOB-2 & \textbf{120.77} & 230.59 & 86.40 (0, 1.172) & 0.0115 \\
\hline
\end{tabular}
\label{table:results_a}
\end{table}

\begin{table}[h]
\caption{$Z_b$ metrics for the experimental data.}
\centering
\begin{tabular}{|c|c|c|c|}
\hline
Controller & Bandwidth & Overshoot & Rising Time \\
\hline
DOB-1 & \textbf{17.97 rad/s} & 0.0 & \textbf{0.14 s} \\
DOB-2 & 13.64 rad/s & 0.0 & 0.18 s \\
\hline
\end{tabular}
\label{table:results_b}
\end{table}

These improvements, as revealed by the proposed metrics, are difficult to observe with traditional blocked-actuator tests alone. The metrics offer a more nuanced, quantitative view of controller behavior, capturing relations not easily seen from simple step or frequency responses.

% \fi

\IEEEpeerreviewmaketitle

 \section{Conclusion}
This short paper introduced a set of quantitative metrics for evaluating force/torque controllers, with a particular focus on assessing how variations in load affect the performance and stability of torque-controlled actuators. The proposed framework includes four metrics: the Load Influence Index (LII), Transparency Residual (TR), Passivity Index Interval (PII), and Load Robustness Threshold (LRT). These metrics provide a structured and interpretable means of characterizing controller behavior under changing load conditions, enabling systematic comparison during controller design and evaluation.

We demonstrated the utility of these metrics through experiments involving a linear actuator driven by different configurations of disturbance observer (DOB)-based controllers. Identified models from system identification were used to compute the metrics and revealed differences in transparency and robustness between controller tunings, despite their similar frequencies responses.

Traditional qualitative criteria offer only limited insight when transparency curves intersect or differ across frequency bands. In contrast, the proposed metrics synthesize complex frequency-dependent behavior into scalar values that are easier to interpret and compare. 
The results support the adoption of these metrics as practical tools for analyzing interaction dynamics and guiding the design of robust and transparent mechatronic systems. This contributes to the development of safer and more effective torque-controlled actuators for physical interaction. Future research may explore applications to a broader range of actuators, generalization to nonlinear systems, and direct computation from frequency response data.

\section*{Acknowledgment}

We thank the Altair Robotics Lab of Verona University, the Robotic Systems Lab of ETH Zurich, and the Legged Robotics Group of thr University of São Paulo for their support.
This research has received funding from the São Paulo Research Foundation (FAPESP) grants 2018/15472-9, 2020/12730-7, 2021/03373-9, 2021/15179-2, and 2022/05048-0.



\bibliographystyle{IEEEtran}
\bibliography{references}

\end{document}